\documentclass[aps,prx,reprint,superscriptaddress]{revtex4-2}
\usepackage{amsmath,amsfonts,amssymb}
\usepackage{graphicx,float,calc,bm}
\usepackage[breaklinks,colorlinks,urlcolor=blue,citecolor=blue,linkcolor=blue]{hyperref}

\begin{document}

\title{Strongly-tilted field induced Hamiltonian dimerization and nested quantum scars in the 1D spinless Fermi-Hubbard model}

\author{Wei-Jie Huang}
\affiliation{CAS Key Laboratory of Quantum Information, University of Science and Technology of China, Hefei, Anhui 230026, China}
\affiliation{CAS Center For Excellence in Quantum Information and Quantum Physics, University of Science and Technology of China, Hefei, Anhui 230026, China}
\affiliation{Hefei National Laboratory, University of Science and Technology of China, Hefei Anhui 230088, China}

\author{Yu-Biao Wu} \email{wuyb@iphy.ac.cn}
\affiliation{Beijing National Laboratory for Condensed Matter Physics, Institute of Physics, Chinese Academy of Sciences, Beijing 100190, China}

\author{Guang-Can Guo}
\affiliation{CAS Key Laboratory of Quantum Information, University of Science and Technology of China, Hefei, Anhui 230026, China}
\affiliation{CAS Center For Excellence in Quantum Information and Quantum Physics, University of Science and Technology of China, Hefei, Anhui 230026, China}
\affiliation{Hefei National Laboratory, University of Science and Technology of China, Hefei Anhui 230088, China}

\author{Wu-Ming Liu}
\affiliation{Beijing National Laboratory for Condensed Matter Physics, Institute of Physics, Chinese Academy of Sciences, Beijing 100190, China}

\author{Xu-Bo Zou} \email{xbz@ustc.edu.cn}
\affiliation{CAS Key Laboratory of Quantum Information, University of Science and Technology of China, Hefei, Anhui 230026, China}
\affiliation{CAS Center For Excellence in Quantum Information and Quantum Physics, University of Science and Technology of China, Hefei, Anhui 230026, China}
\affiliation{Hefei National Laboratory, University of Science and Technology of China, Hefei Anhui 230088, China}

\begin{abstract}
We investigate the quantum dynamics of the 1D spinless Fermi-Hubbard model with a linear-tilted potential.
Surprisingly in a strong resonance regime, we show that the model can be described by the kinetically constrained effective Hamiltonian,
and it can be spontaneously divided into two commuting parts dubbed Hamiltonian dimerization,
which consist of a sum of constrained two-site hopping terms acting on odd or even bonds.
Specifically it is showed that each part can be independently mapped onto the well-known PXP model,
therefore the dimerized Hamiltonian is equivalent to a two-fold PXP model.
As a consequence, we numerically demonstrate this system can host the so-called quantum many-body scars, which present persistent dynamical revivals and ergodicity-breaking behaviors. 
However in sharp contrast with traditional quantum many-body scars, here the scarring states in our model driven by different parts of Hamiltonian will oscillate in different periods,
and those of double parts can display a biperiodic oscillation pattern,
both originating from the Hamiltonian dimerization.
Besides, the condition of off-resonance is also discussed and we show the crossover from quantum many-body scar to ergodicity breaking utilizing level statistics. 
Our model provides a platform for understanding the interplay of Hilbert space fragmentation and the constrained quantum systems.
\end{abstract}

\maketitle

\section{Introduction}
The rapid experimental progress of recent years in the preparation and controllability of quantum coherent systems has enabled the realization of quantum many-body systems in well-isolated settings and the investigation of their nonequilibrium dynamics, mainly utilizing cold atoms
\cite{Kinoshita2006nature,Schreiber2015science,Kaufman2016science,Choi2016science}, 
trapped ions
\cite{Smith2016np,Kaplan2020prl,Schindler2013np}, 
and superconducting circuits
\cite{Chiaro2022prr,Fitzpatrick2017prx,Guo2019prappl,Ye2019prl}.
The most common many-body systems reach their thermal equilibrium under long unitary time evolution, which can be explained by the eigenstate thermalization hypothesis (ETH) \cite{Deutsch1991pra,Srednicki1994pre,Srednicki1999jpa}.
It states that the expectation value of local observables only depends on the eigenstates, behaving as a basic thermal ensemble for quantum ergodicity.
Most quantum many-body systems are known to obey ETH.
Whereas, some systems have been proposed that violate ETH,
for instance, exactly solvable systems with fixed parameters
or many-body localization (MBL) phases with strong disorders
\cite{Gornyi2005prl,Basko2006ap,Nandkishore2015arcmp,Abanin2017ap,Alet2018crp,Abanin2019rmp,Serbyn2013prl}.

More recently, another new mechanism of ETH violation called quantum many-body scar (QMBS)\cite{Ho2019prl,Turner2021prx,Bluvstein2021sci,Turner2018prb,Lin2020prr,Mondragon-Shem2020arx,Zhao2020prl,Iadecola2019prb,Chattopadhyay2020prb,Mukherjee2020prb}
has been reported in a Rydberg atom quantum simulator, where the majority of the initial states reach the corresponding thermal ensembles, while some special initial states show long-lived oscillation \cite{Bernien2017nature}.
Soon afterward, this system was pointed out theoretically to be described by the so-called PXP model
\cite{Turner2018np,Choi2019prl,Khemani2019prb,Michailidis2020prr,Lin2020prb}.
And various theoretical explanations for the underlying mechanism have been proposed to investigate the ubiquity of QMBS
\cite{Bull2019prl,Kuno2020prb,Serbyn2021np,Moudgalya2020prb,Pakrouski2020prl,O'Dea2020prl,Surace2020qua}.
Later, QMBS has also been found to exist in other kinds of models,
such as the Affleck-Kennedy-Lieb-Tasaki (AKLT) model and the spin-1 XY model
\cite{Moudgalya2018prb,Moudgalya2020prb,Mark2020prb,Schecter2019prl,Chattopadhyay2020prb}.

On the other hand, it has been proposed that ergodicity breaking can occur due to the Hilbert space fragmented into exponentially many disconnected subspaces,
which typically results from the presence of constrained hopping terms and higher-moment symmetry, such as the dipole conservation
\cite{Sala2020prx,Moudgalya2021arxiv,Khemani2020prb,Yang2020prl}.
Furthermore, Hilbert space fragmentation as well as the induced QMBS have been investigated and observed in the spinful fermionic and hard-core bosonic Hubbard models with an applied gradient field 
\cite{Scherg2021nc,Desaules2021prl}.
In these models, the constrained hopping terms of two consecutive sites are responsible for rendering the ergodicity breaking.
An immediate question is natural to come up whether novel phenomena can emerge in the systems with the constrained hopping terms of longer-range sites.

In this Letter, we investigate the nonergodic behaviors in the 1D strongly-tilted spinless Fermi-Hubbard model, which exhibits QMBS with quite a few unique features compared with the aforementioned models.
We show that the kinetically constrained effective Hamiltonian can be spontaneously divided into two commuting parts acting on four consecutive sites, 
which can be divided into the isolated odd and even part of Hamiltonian, dubbed Hamiltonian dimerization.
Each isolated part of the Hamiltonian can be mapped to the so-called PXP model by considering all allowed four-site configurations with those three-site configurations in PXP model.
Therefore each part can independently give rise to QMBS, showing long-lived period oscillations and ergodicity-breaking many-body eigenstates.
As for the initial states that can be acted under both two parts of the Hamiltonian, there exist special eigenstates exhibiting similar behaviors as QMBS. 
Nevertheless, the periodic oscillations depend on both the odd and even part of Hamiltonian.
In particular, if the two individual periods have a considerable difference in size, 
additional persistent revivals over one large period will emerge.
These unique phenomena can be explained by the two isolated part of Hamiltonian contributing to the two-fold PXP model and in turn, verifying the Hamiltonian dimerization.
We further show that the scarred states rapidly disappear when the tilted potential presents finite detuning with the interaction strength.

The remainder of the paper is organized as follows.
In Sec. \ref{section_dimerization}, we introduce the Hamiltonian we study and new concept called Hamiltonian dimerization.
In Sec. \ref{section_single_scar}, we investigate the QMBS under single part of Hamiltonian.
In Sec. \ref{section_nested_scar}, we further explore the whole part of Hamiltonian, while the fidelity dynamics shows period-renormalized oscillations and the nested pattern. Then we called this phenomenon as nested scar and explain it by use of nested energy separation in the state overlap.
In Sec. \ref{section_off_resonant}, we consider the Hamiltonian in the off-resonance regime and argue that the QMBS will be broken by increasing the strength of the off-resonant.
In Sec. \ref{section_level_statistics}, we prove that the Hamiltonian presents two different level statistics at the same conditions based on the specific initial states.
We argue the experimental feasibility in Sec. \ref{section_experimental} and conclude in Sec. \ref{section_conclusion} with a conclusion and discussion.
The Appendixes provide further discussion about the effect of off-resonance on the single part of Hamiltonian and the whole part of Hamiltonian. Appendixes \ref{section_pxp_model} shows the mapping between tilted Fermi-Hubbard model and double PXP model.

\section{Hamiltonian dimerization} \label{section_dimerization}
We start from the 1D tilted spinless Fermi-Hubbard model governed by the Hamiltonian

\begin{align}
	H = &-\frac{J_{\rm o}}{2}\sum_{j}c_{2j+1}^{\dag}c_{2j+2}-\frac{J_{\rm e}}{2}\sum_{j}c_{2j}^{\dag}c_{2j+1}+{\rm H.c.}	\notag\\
		   &+\sum_{j=1}^{L}\Delta j(n_{j}-\frac{1}{2})+U\sum_{j=1}^{L-1}(n_{j}-\frac{1}{2})(n_{j+1}-\frac{1}{2})
		   \,.	\label{eq-h}
\end{align}

\begin{figure}[t]
\centering
\includegraphics[width=0.45\textwidth]{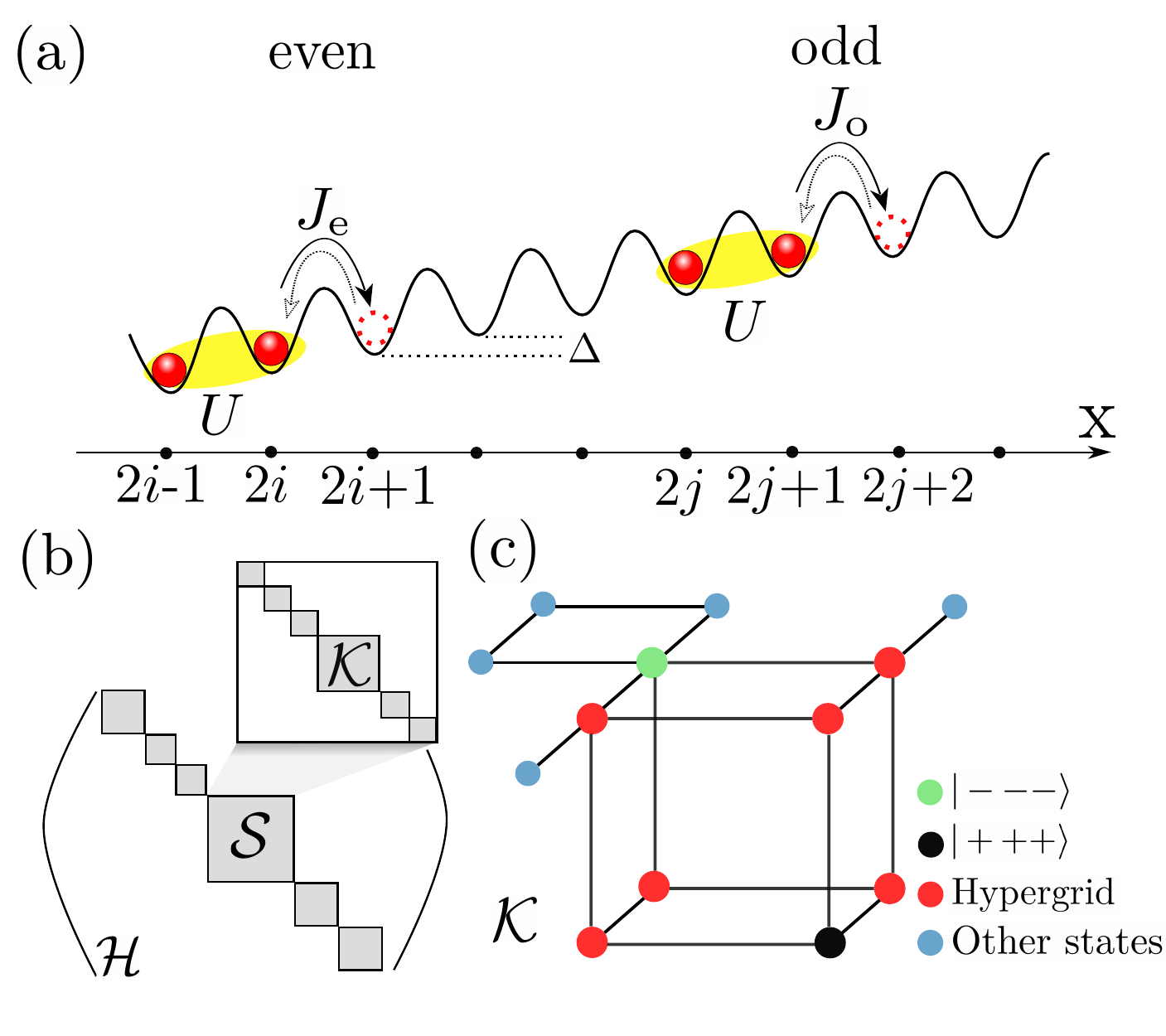}
\caption{(a) Schematic of the physical system, $\Delta$ is the on-site potential energy, $J_{\rm o} (J_{\rm e})$ is the hopping terms acting on odd (even) bonds and $U$ is the interaction terms. Here ..., 2i-1, 2i, 2i+1, ... stands for the sites on the lattice.
(b) The graph respresents the illustration of Hilbert-space fragmentation. The symmetry sectors $\mathcal{S}$ of the total Hilbert space $\mathcal{H}$ decouple into disconnected fragments $\mathcal{K}$.
(c) The restricted Hibert space graph of the effective model (the largest $\mathcal{K}$) for $L=12$. Red vertices denote the states belonging to the hypergrid, with the black (green) vertices corresponding to $|+++\rangle$ ($|---\rangle$) state defined in the text. For this graph, the hypergrid contains 8 vertices out of 13. }
\label{fig.model}
\end{figure}

Here $L$ is the system size. 
$c_{j}^{\dag}$ $(c_j)$ denotes the electron creation (annihilation) operator at site $j$ with the number operator $n_{j}=c_{j}^{\dag}c_{j}$. 
$J_{\rm o}$ $(J_{\rm e})$ is the strength of hopping that acts on the odd(even) bonds, 
and $U$ denotes the nearest-neighbor interaction strength.
The tilted potential $\Delta$ can be generated by applying a magnetic field gradient along the optical lattice as shown in Fig. \ref{fig.model}(a), which introduces a dipole term.
In the following, the calculations are performed in open boundary conditions.
We firstly focus on the strong-tilted field and resonance regime $\Delta=U \gg J_{\rm o},J_{\rm e}$.
In the rotating frame (Appendix \ref{section_effective_model}), the Hamiltonian reduces to $H_{\rm eff} = H_{\rm odd}+H_{\rm even}$ with
\begin{align}
	H_{\rm odd}	= &-\frac{J_{\rm o}}{2}\sum_{j}n_{2j}c_{2j+1}^{\dag}c_{2j+2}(1-n_{2j+3})+ {\rm H.c.} \,,   \notag\\ 
	H_{\rm even}= & -\frac{J_{\rm e}}{2}\sum_{j}n_{2j-1}c_{2j}^{\dag}c_{2j+1}(1-n_{2j+2})+ {\rm H.c.}    \,, 
	\label{eq-h-eff}
\end{align}
in which $H_{\rm odd}$ and $H_{\rm even}$ denotes the constrained hopping terms acting on odd or even bonds as shown in Fig. \ref{fig.model}(a).
The hopping between the two middle sites is only permitted when the left surrounded site is occupied and the right surrounded site is empty.
In what follows, we use $|1\rangle$ to refer the occupied state and $|0\rangle$ refer to the vaccum state.
Here ${\rm H.c.}$ stands for the Hermitian conjugation.

The effective Hamiltonian (\ref{eq-h-eff}) has two conserved quantities (see Appendix \ref{section_fragmentation}), which organize the Hilbert space $\mathcal{H}$ into small symmetry sectors $\mathcal{S}$. Because of the hopping process constraint, the Hilbert space on the symmetry sector can be further fractured into many disconnected subsectors $\mathcal{K}$. The corresponding process of Hilbert space fragmentation is shown in Fig. \ref{fig.model}(b).
The connected component of the Hilbert space only allows certain four-site configurations 
($|1100\rangle$, $|1010\rangle$), while all other four-site configurations are forbidden.
It is noted that both the two allowed four-site configurations only support the hopping process of the middle two sites, whereas they cannot generate the hopping process of the left or right one site whatever kind of configuration is added into the ends. 
Therefore, the total effective Hamiltonian can be spontaneously divided into two isolated parts, 
i.e. $H_{\rm odd}$ and $H_{\rm even}$, according to the contained hopping terms acting on odd or even bonds, as shown in Fig. \ref{fig.model}(a).
And thus we call it Hamiltonian dimerization.

With respect to the Hilbert space, the dimerized Hamiltonians that commute among each other, i.e. $[H_{\rm odd},H_{\rm even}]=0$, give rise to different oscillating behaviours.
In Table. \ref{tab-class}, we list four different scenarios.
The first two situations represent the states can only act under one isolated part,
and type three is for those that can act under both $H_{\rm odd}$ and $H_{\rm even}$.
The last one does not contain the permitted modes for hopping, so states in this situation cannot evolve under the total effective Hamiltonian.
Below we discuss the difference of persistent revivals under isolated part and double parts.

\begin{table}[!htbp]
	\tabcolsep = 22 pt
	\centering 
	\caption{Hilbert space classification under the containing Hamiltonian,  $H_{\rm odd}$ or $H_{\rm even}$,
	 according to the Hamiltonain dimerization.
	 And we give the example states for different parts. }
	\begin{tabular}{ccc}
		\hline\hline
		$H_{\rm odd}$ & $H_{\rm even}$ & Example State \\
		\hline
		\checkmark & $\times$ &  $|01100110\cdots\rangle$  \\
		$\times$ & \checkmark &  $|11001100\cdots\rangle$  \\
		\checkmark & \checkmark & $|110001100\cdots\rangle$  \\
		$\times$ & $\times$ & $|11101110\cdots\rangle$  \\
		\hline\hline
	\end{tabular}
	\label{tab-class}
\end{table}

\section{Single scar under resonant regime} \label{section_single_scar}
The restricted Hilbert space graph has been used to analyse the dynamic structure.
Because the evolution of system depends on four adjacency sites, we take four adjacency sites as a cell.
Each cell can take the values $|+\rangle=|1100\rangle$ and $|-\rangle=|1010\rangle$.
We focus on the largest connected component, which is the one containing the $|\mathbb{Z}_2\rangle$ state.
In Fig. \ref{fig.model}(c), we plot the restricted Hilbert space (the largest $\mathcal{K}$) of the effective model for $L=12$.
Each vertice stands for a basis state.
Two vertices are connected by the nonzero matrix element between the two corresponding basis states.
Basis states hopping inside cells constitute a form of the hypergrid, which occupies eight vertices of the cube.
On the other hand, basis states hopping between cells are outside the hypergrid, that are labelled by blue.
As we will show below, the black vertice $|110011001100\rangle$ ($|+++\rangle$) without edges out of the hypergrid hosts persistent oscillation,
and the green vertice $|101010101010\rangle$ ($|---\rangle$) that is connected by the blue vertices thermalizes.
Taking consideration of intercommunity of odd part and even part,
in what follows we calculate the dynamic properties and eigenstates of only isolated even part of Hamiltonian to further investigate the system with exact diagonalization (ED)\cite{Sandvik2010aip}.

\begin{figure}[t]
\centering
\includegraphics[width=0.49\textwidth]{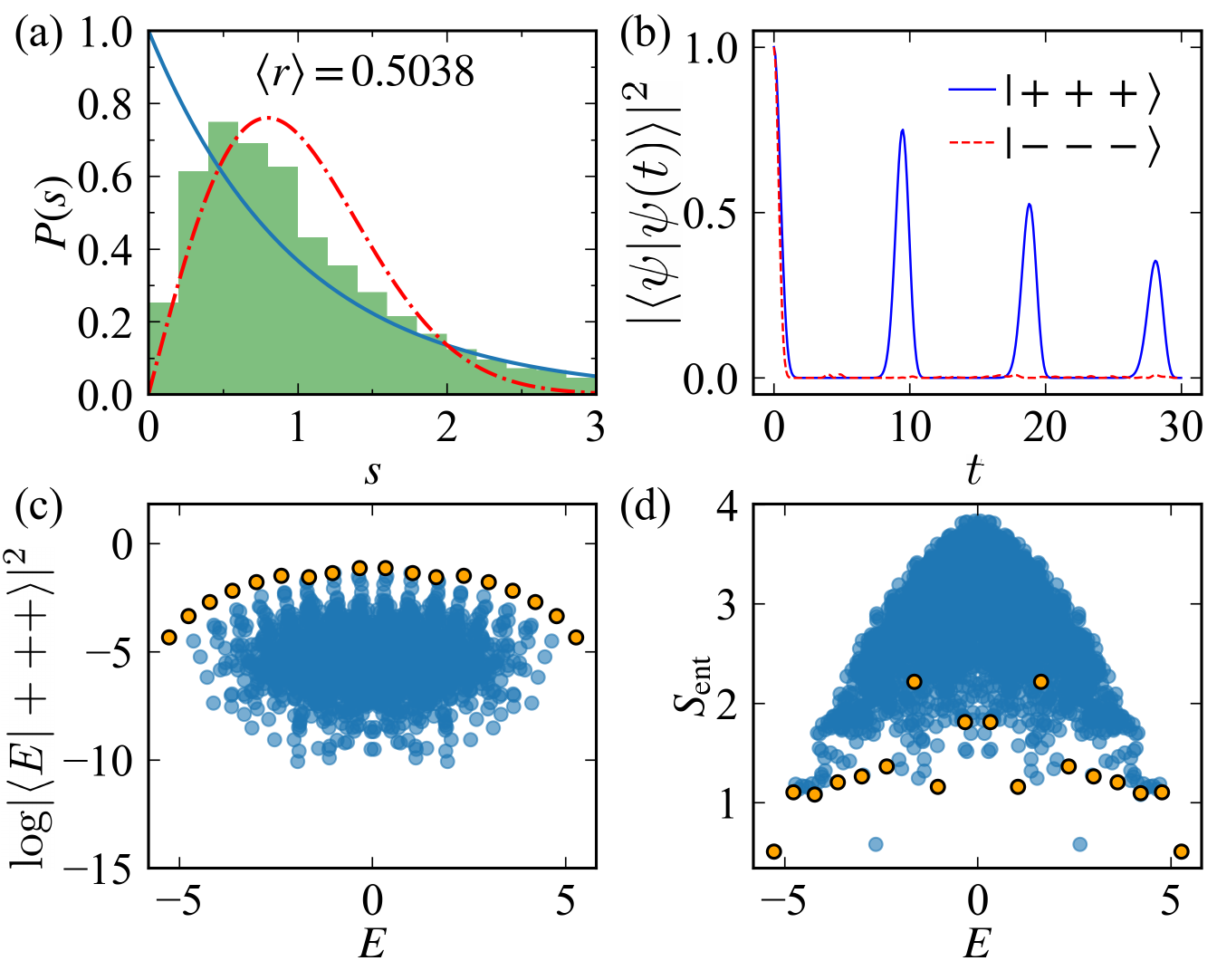}
\caption{(a) The level spacing distribution with $L=44$. The blue solid (red dashdot) curve denotes Possion distributions (Wigner-Dyson distributions). The $\langle r\rangle$ value of the distribution given in the figure is close to the Wigner-Dyson result, $\langle r\rangle\approx 0.5038$. 
(b) Fidelity dynamics with $L=36$ for $|+++\rangle$ and $|---\rangle$ state. 
(c)  Overlap between eigenstates and $|+++\rangle$ state as a function of corresponding energy $E$. 
(d) Entanglement entropy $S_{\rm ent}$. The red dots indicate the eigenstates sitting at the top of each tower of states. All other parameters are: system size is $L=36$ and the hopping strength is $J_{\rm e}=J_{\rm o}=1$.}\label{fig.distribution_fix}
\end{figure}

The Hamiltonian possesses a symmetry $F$ related to spatial inversion and particle-hole exchange having eigenvalues $\pm 1$.
It can be expressed by $F=\Pi_{j=1}^{L/2}F_{j}$,
here $F_{j}=n_{j}(1-n_{L+1-j})+(1-n_{j})n_{L+1-j}+c_{j}^{\dag}X_{j,L-j}c_{L+1-j}^{\dag}+c_{L+1-j}X_{j,L-j}c_{j}$,
with $X_{j,L-j}=(2n_{j}-1)(2n_{j+1}-1)\cdots(2n_{L-j}-1)$.
Furthermore, the model has another symmetry $M=e^{i\pi\sum_{j}n_{2j+1}}$ anticommutes with the Hamiltonian, and therefore each eigenstate $|\psi\rangle$ with energy $E\ne 0$ has a partner $M|\psi\rangle$ with energy $-E$.
Here we calculate the average ratio of consecutive level spacing and corresponding level spacing distribution in Fig. \ref{fig.distribution_fix}(a).
We find the level spacing distribution $P(s)$ follows the Wigner-Dyson distribution, 
while $\langle r\rangle=0.5038$ is consistent with the Gaussian orthogonal ensemble\cite{Buijsman2017prl,Bohigas1984prl,Robnik2016epj,Victor1977prs,Atas2013prl}.
Taken together, these facts suggest that the even part of Hamiltonian Eq. (\ref{eq-h-eff}) is ergodic.

The practical determination of QMBS is that some special initial states manifest initial memory with the development of time, while the majority of initial states thermalize and do not keep initial memory.
Fidelity dynamics $|\langle\psi|\psi(t)$ $\rangle|^{2}$ is a proper quantity to detect QMBS in 
Fig. \ref{fig.distribution_fix}(b) for even part of Hamiltonian.
As we mentioned above, periodic oscillations are clearly visible in $|+++\rangle$ state with the period $T\approx 10$.
The initial state $|+++\rangle$ is in the extreme corner of the restricted Hilbert space and it disconnected with the states out of the hypergrid.
The disconnected links with other states in the hypergrid make the initial state show revivals.
On the contrary, the $|---\rangle$ state does not oscillate and the initial information is lost, which contributes to the high connectivity with the states out of the hypergrid.
Apart from the fidelity calculation, we can also verify QMBS by means of mapping to the PXP model
(see Appendix \ref{section_pxp_model}).

To provide further insights into the structure of special eigenstates, we study the entire many-body spectrum according to the overlap with special eigenstates.
Fig. \ref{fig.distribution_fix}(c) illustrates the scatter plot of the overlap $|\langle E|+++\rangle|^{2}$ between the initial state $|+++\rangle$ and eigenstates $|E\rangle$ on a log scale versus the eigenenergy.
The majority of eigenstates huddle together, while some special states are distinguished by atypically high overlaps with the $|+++\rangle$ state.
The energy separation between the special states stays approximately constant near zero energy 
and is given by 0.66,
thus the revival period is $T=9.52$ which is consistent with the estimated period above.

It is well known that for highly excited ETH eigenstates, the entanglement entropy obeys the volume law, scaling proportionally to the volume\cite{Bianchi2022prxq}.
As we will see, entanglement properties of QMBS special eigenstates are dramatically different.
Fig. \ref{fig.distribution_fix}(d) shows the entanglement entropy of the entire eigenstates.
The majority of eigenstates are consistent with the thermalization assumption.
However, those special states labelled by the red dots which have high overlaps with the initial state exhibit apparently low entanglement entropy, causing an ETH violation.

\section{Nested scar under resonant regime}   \label{section_nested_scar}
In contrast to the investigations on QMBS under isolated part of Hamiltonian, those under double parts of Hamiltonian need considering Hilbert space fragmentation in two independent subspaces.
In this case, initial states play a important role in the evolution and ergodicity.
Specifically speaking, evolutions only take place in odd (even) part of Hilbert spaces if some parts of initial states can match those acting under odd (even) part of Hamiltonian.
Based on this understanding, the initial state $|\psi_{\rm dimer}\rangle$ is set as the combination of odd and even parts: $|\psi_{\rm dimer}\rangle$=$|11001100\rangle\otimes |0\rangle \otimes |11001100\rangle$ ($|++0++\rangle$).
$|11001100\rangle$ ($|++\rangle$) is exactly enclosed vertice of the even Hilbert subspace which shows revivals in 
Fig. \ref{fig.distribution_fix}(b).
Interestingly, the initial state with middle additional site of empty occupancy which introduce the combination of double parts will exhibit unique oscillating behaviours.

\begin{figure}[t]
\centering
\includegraphics[width=0.49\textwidth]{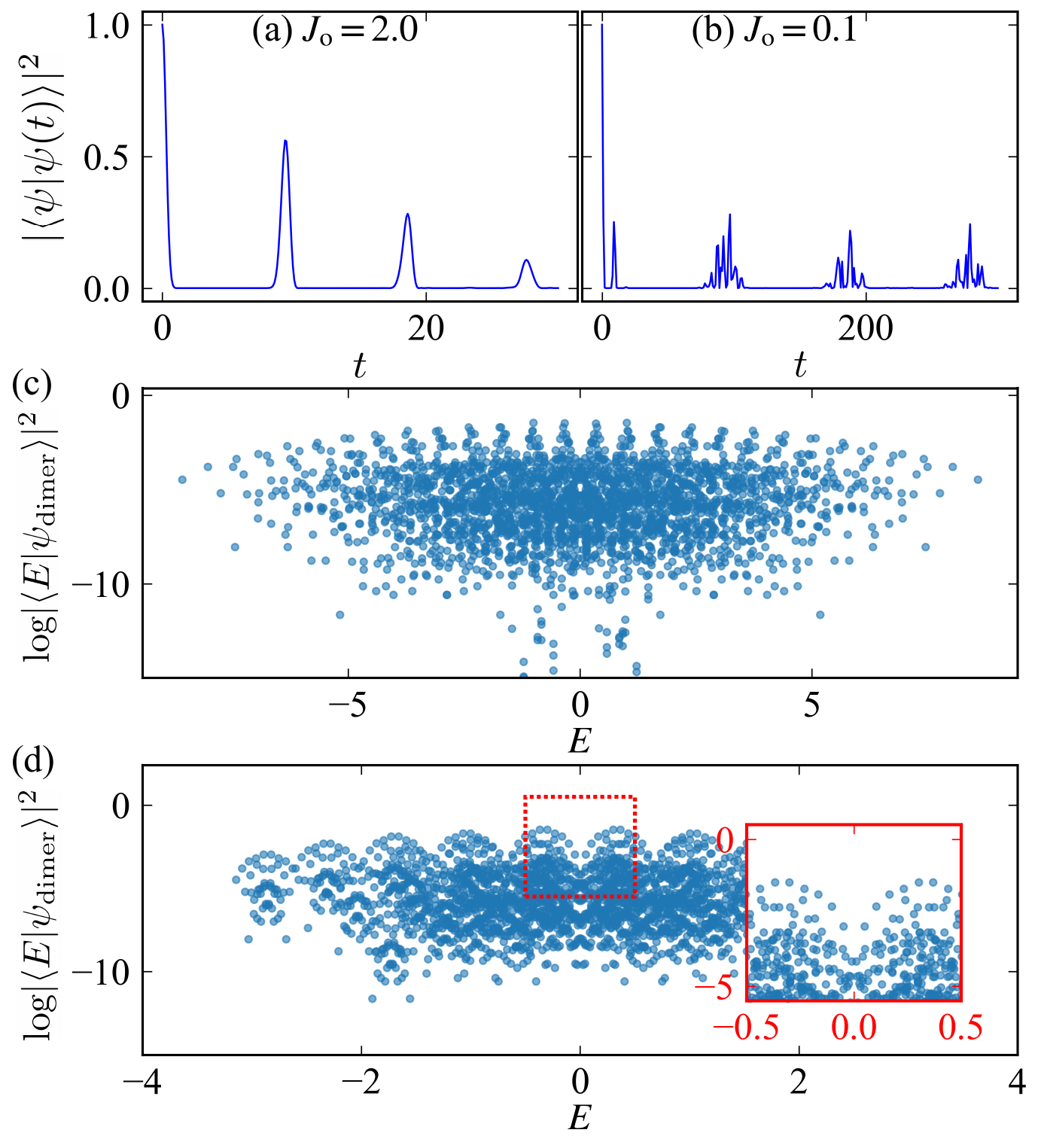}
\caption{Fidelity dynamics with $L=41$ for $|\psi_{\rm dimer}\rangle$ defined in text and $J_{\rm e}=1$ at odd sites hopping (a) $J_{\rm o}=2.0$ and (b) $J_{\rm o}=0.1$. The overlap with $J_{\rm e}=1.0$,  (c) $J_{\rm o}=2.0$ and  (d) $J_{\rm o}=0.1$ between eigenstates and combined initial state $|\it{\Psi}_{{\rm dimer}}\rangle$ as a function of corresponding energy $E$.  The inset shows energies around zero at $J_{\rm e}=1,J_{\rm o}=0.1$. The system size is $L=41$. }\label{fig.fidelity_both}
\end{figure}

For simplicity without loss of generality,
we set the hopping of the even part $J_{\rm e}=1.0$ as the energy unit, and change the hopping of the odd part $J_{\rm o}$ for different results.
Fig. \ref{fig.fidelity_both}(a,b) show the fidelity dynamics for combined initial state $|\psi_{\rm dimer}\rangle$.
As we can see, fidelity keeps oscillations in $J_{\rm o}=2.0$ and $J_{\rm o}=0.1$ both,
which indicates the dimerized Hamiltonian is nonergodic.
Since the combined initial state hosts persistent oscillation under either isolated odd and even part Hamiltonian,
the total evolution still shows nonergodic.
For $J_{\rm o}=2.0$ in Fig. \ref{fig.fidelity_both}(a), the dynamics of fidelity shows clear oscillations of period $T_{\rm dimer}\approx 10$.
While for $J_{\rm o}=0.1$ in Fig. \ref{fig.fidelity_both}(b), the oscillation period increases approximately ten times.
It can be interpreted by the commutation of two dimerized Hamiltonian parts, 
i.e.  $[H_{\rm odd},H_{\rm even}]=0$,
which ensures that the evolution of the isolated part of Hamiltonian is independent with the other one,
giving rise to period-renormalized oscillations and the nested pattern.
In other words, the total period of fidelity dynamics in the dimerized Hamiltonian is the lowest common multiple of the period of fidelity dynamics in the isolated part of Hamiltonian.
For example, $J_{\rm o}=2.0$ with a halve period makes no difference to the total period compared with Fig. \ref{fig.distribution_fix}(b).
while $J_{\rm o}=0.1$ renormalizes the total period to ten times, 
Moreover, the amplitudes of revivals under double parts of Hamiltonian are obviously less than those under isolated part of Hamiltonian, because the total amplitudes of revivals come from the production of two isolated parts.

As shown in Fig. \ref{fig.fidelity_both}(b), there are additional oscillations consistent to the original period in every revival. 
It can be explained by the overlap between eigenstates and special initial state.
In Fig. \ref{fig.fidelity_both}(c), the large overlap states also form a band on top of all states, 
and $J_{\rm o}=2.0$ makes no change to the energy separation,
thus the total period keep consistent with that in the former text.
However for $J_{\rm o}=0.1$ in Fig. ~\ref{fig.fidelity_both}(d), 
the large overlap states form a set of towers on top of all states,
and every tower consists of several points with equal energy spacing.
After comparing the energy separation in details, we find that the energy separation of each tower is 0.66,
which is consistent with the uniform regime,
and the inner energy spacing is around 0.06, contributing the large-period oscillations.
In brief, the nested QMBS band in the two-fold PXP model is tunable in a relative regime due to the commutation of two dimerized parts, and it can provide a better understanding of Hamltonian dimerization.

\section{Scar under off-resonant regime}	\label{section_off_resonant}
Above we have discussed the properties of the Hamiltonian under the resonance regime,
the it is natural to extend the off-resonance regime.
The off-resonance term of Hamiltonian is written as

\begin{align}
	H_{\rm off} =      \delta\sum_{j}(n_{j}-\frac{1}{2})(n_{j+1}-\frac{1}{2})
			             \,, \label{eq-h-off}
\end{align}

where $\delta=U-\Delta$ denotes the strength of the off-resonance interaction.

Because the off-resonance interaction only refer to the Hamiltonian diagonal matrix element,
the connections between different basis states are not changed,
i.e. the Hilbert space fragmentation is in accordance with it at resonance regime.
Though the off-resonance term does not commute with the odd or the even part of Hamiltonian,
the commutation relation between odd and even part of Hamiltonian stays the Hamiltonian dimerization.

\begin{figure}[t]
\centering
\includegraphics[width=0.49\textwidth]{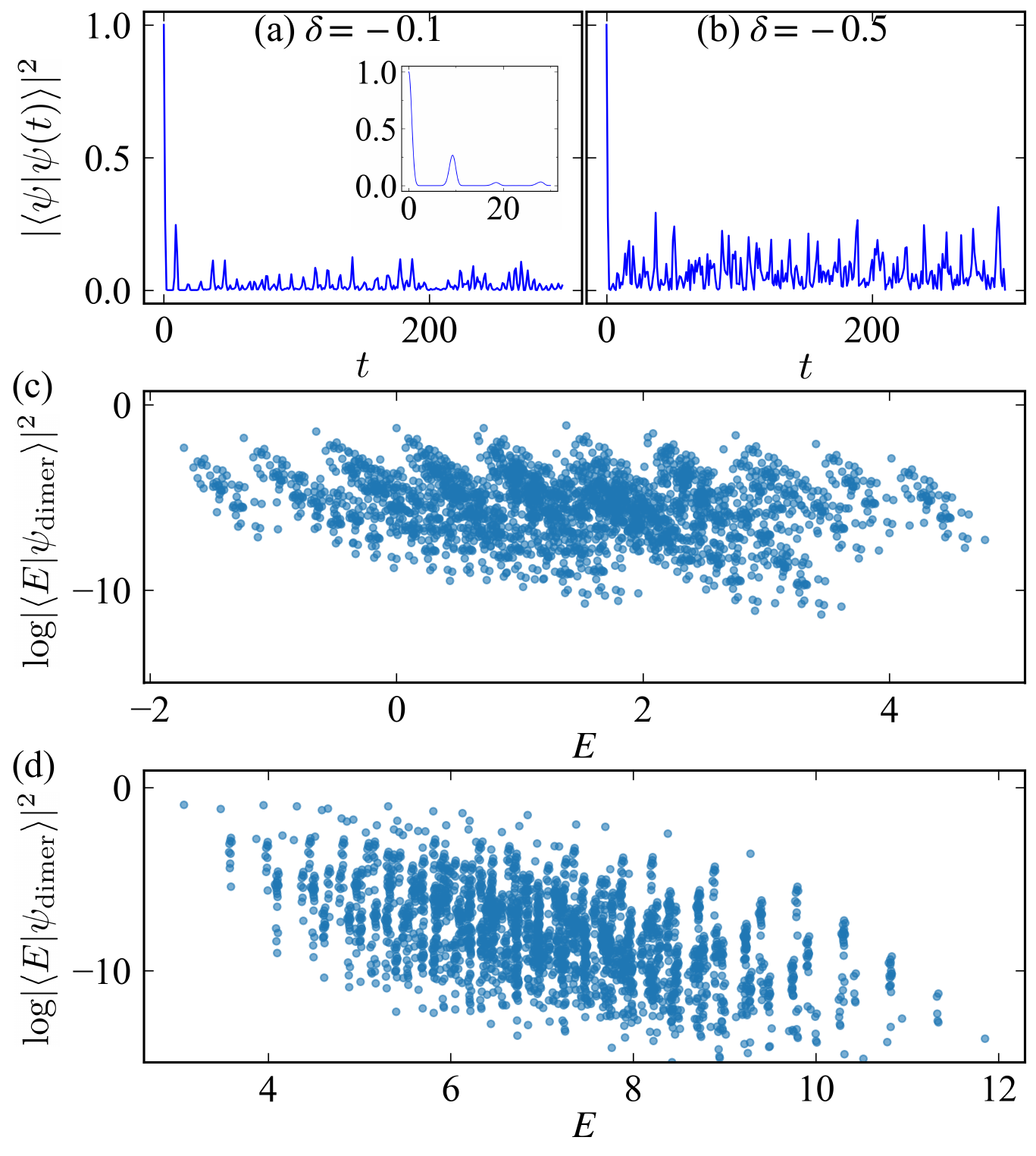}
\caption{Fidelity dynamics with $L=41$ for $|\psi_{\rm dimer}\rangle$ defined in text and $J_{\rm e}=1$, $J_{\rm o}=0.1$ at off-resonance regime (a) $\delta=-0.1$ and (b) $\delta=-0.5$. The inset shows fidelity dynamics in time intervel [0, 30]. The overlap with $J_{\rm e}=1.0$, $J_{\rm o}=0.1$ (c) $\delta=-0.1$ and  (d) $\delta=-0.5$ between eigenstates and combined initial state $|\it{\Psi}_{{\rm dimer}}\rangle$ as a function of corresponding energy $E$. The system size is $L=41$. }\label{fig.fidelity_both_off}
\end{figure}

The effect of the off-resonance on the nested scar is illustrated in Fig. ~\ref{fig.fidelity_both_off}(a,b) which plots the fidelity dynamics of combined initial state $|\psi_{\rm dimer}\rangle$ in $J_{\rm e}=1.0$ and $J_{\rm o}=0.1$ with off-resonance strengths $\delta=-0.1,-0.5$.
For $\delta=-0.1$, we observe that the period-renormalized oscillations and the nested pattern vanish,
compared with Fig. ~\ref{fig.fidelity_both}(b).
Specially, the fidelity dynamics show usual periodic oscillations as single scar when we focus on the time intervel [0, 30].
For $\delta=-0.5$, system presents neither the property of single scar nor nested scar because of the irregular oscillation.
However the fact that the overall value of time evolution is larger than it in resonance regime indicates ergodicity breaking.

In constrast with the overlap between eigenstates and combined initial state at resonance regime,
Fig. ~\ref{fig.fidelity_both_off}(c, d) plots the overlap at off-resonance regime.
In Fig. ~\ref{fig.fidelity_both_off}(c), the large overlap states is much better sperated from the bulk with the energy separation 0.66,
which is consistent with the above period of fidelity dynamics $T\approx 10$.
However each tower of states is not comprised of the additional large overlap state,
leading to the vanishment of the period-renormalized oscillations.
In Fig. ~\ref{fig.fidelity_both_off}(d), the large overlap states with equal energy separation is inexistence and all of state huddle together.

The fidelity dynamics and overlap between eigenstates and initial states with the other off-resonance interaction( $\delta=-0.01, -4.0$) is shown in Fig. ~\ref{fig.off_resonance_both}.
The dynamics present the nested scar (ergodicity breaking) at weak (strong) off-resonance interaction.
Actually, the transition from nested scar to single scar and finally ergodicity breaking by increasing off-resonance interaction can be explained by the Hamiltonian dimerization.
When the off-resonance interaction is weak ($\delta=-0.01$), both of odd and even part of Hamiltonian is still QMBS.
By increasing $\delta=-0.1$, the even part of Hamiltonian is QMBS because the influence of hopping term ($J_{\rm e}=1.0$) is very stronger than the off-resonance term ($\delta=-0.1$).
In constrast, QMBS in the odd part of Hamiltonian is broken because the influence of hopping term ($J_{\rm o}=0.1$) is not stronger than the off-resonance term.
As a result, Hamiltonian dimerization only present single scar.
Finally, the strong off-resonance interaction ($\delta=-4.0$) break QMBS not only in even part but also in odd part of Hamiltonian,
leading to ergodicity breaking. 
In essence, ergodicity breaking is based on the space fragmentationn under the strong off-resonance interaction.
Because the strength of off-resonance interaction is stronger than the strength of hopping, the effect of the hopping is negligible.
The Hilbert space $\mathcal{K}$ is approximately divided into different subspaces according to different states acting on the off-interaction.

\section{Level statistics}	\label{section_level_statistics}
The reason of the crossover from nested scar to single scar indicates that the system presents two different level statistics at the same conditions based on the specific initial states.
We calculate the average energy level spacing ratio within a connected subspace built from the root configuration $|+++\rangle$, which only refers to the even part of Hamiltonian (We set $J_{\rm e}=1.0$, $J_{\rm o}=0.1$, $\delta=-0.5$, $L=40$).
We find $\langle r\rangle \approx 0.49$, consistent with the Gaussian orthogonal ensemble.
It suggest the system is ergodic in the specific subspace. 
If the connected subspace built from the root configuration $|0\rangle\otimes|+++\rangle\otimes|0\rangle$, only considering the odd part of the Hamiltonian.
We find $\langle r\rangle \approx 0.41$ indicating Possion ensemble corresponding to nonergodicicity.

\begin{figure}[t]
\centering
\includegraphics[width=0.49\textwidth]{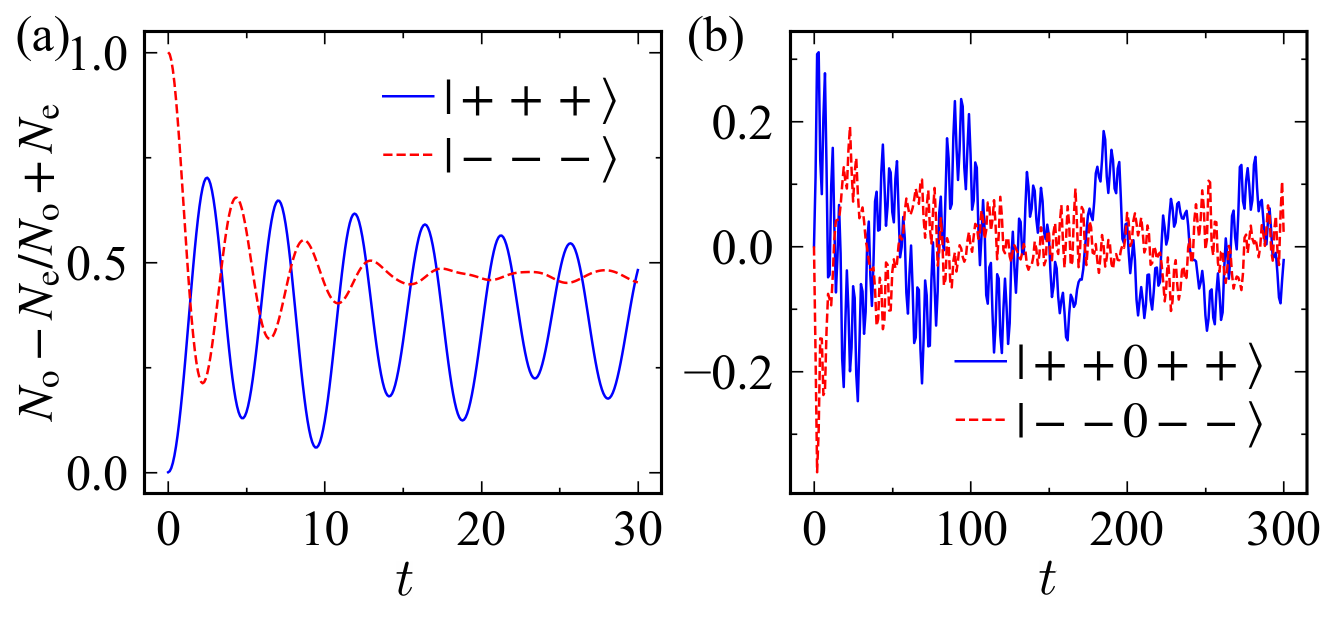}
\caption{(a) The time evolution of the occupancy imbalance with initial state $|+++\rangle$ and $|---\rangle$. The system size is $L=36$, $J_{\rm e}=J_{\rm o}=1.0$. (b) The time evolution of the occupancy imbalance with initial state $|++0++\rangle$ and $|--0--\rangle$. The sysytem size is $L=41$, $J_{\rm e}=1.0$ and $J_{\rm o}=0.1$.}\label{fig.overlap_inset}
\end{figure}

\section{Experimental feasibility}	\label{section_experimental}
From an experimental point of view, we note that all major ingredients of the proposed scheme can be readily realized using the existing techniques of ultracold atoms.
We consider a concrete example, e.g. a single-component ultracold Fermi gas of $^{40}$K atoms,
here we choose the hyperfine state $|F,m_F\rangle=|9/2,-7/2\rangle$ 
due to its $p$-wave interaction has been well implemented in previous experiments
\cite{Regal2003prl,Ticknor2004pra,Gunter2005prl}.
We load the atoms into a 1D optical lattice with the wavelength $\lambda_L=826$ nm and trap depth $V_0=9E_R$.
The recoil energy of such an optical lattice is $E_R=h^2/(2m\lambda_L^2)\approx 7.26$kHz.
Using the technique of $p$-wave Feshbach resonance, we can get a bare $p$-wave interaction strength $g_p=550E_R$.
After the above setups, the system parameters used in main text can be determined as
hopping amplitudes $J_{\rm o}=J_{\rm e}\approx 0.0192E_R$ and nearest-neighbor interaction strength 
$U\approx 0.2265E_R \gg J_{\rm o},J_{\rm e}$.
For different hopping amplitudes, we can synthesize the staggered hopping by applying
a double-well structure to the optical lattice potential or using optical fields, 
which was also widely applied in ultracold atoms
\cite{Li2013natcommun,Greif2013sci,Atala2014natphys,Zhang2017pra}.
The recent emulation of the Hubbard model with tilted magnetic fields $\Delta$ using ultracold atoms trapped in optical lattices \cite{Scherg2021nc,Zhou2022science}
allows us to predict that our physics investigations can be experimentally verified with quantum simulators.
In particular, similar many-body scarring in a Bose-Hubbard model with unit-filling and a linear tilted potential was also recently observed \cite{Su2022arxiv}.

For the convenience of experimental implementation, we calculate the occupancy imbalance $\mathcal{I}=N_{{\rm o}}-N_{{\rm e}}/N_{{\rm o}}+N_{{\rm e}}$ which corresponds to staggered magnetization in this system\cite{Chen2020pra}.
Here, $N_{{\rm o}}$ $(N_{{\rm e}})$ is the total number of fermions on the odd (even) sites.
The density imbalance is one of the feasible quantities that can be experimentally measured with current techniques.
For isolated part of Hamiltonian in Fig. \ref{fig.overlap_inset}(a), most of initial states reach a fixed value rapidly, consistent with thermalization of the system. And the fixed value of thermalization $\approx 0.47$ that is not zero.
By contrast, QMBS remains persistent oscillations of the occupancy imbalance, whose oscillation period matches half the wave function revival period.
For double parts of Hamiltonian in Fig. \ref{fig.overlap_inset}(b), the occupancy imbalance of thermalization state is fixed at zero, and QMBS also shows robust oscillations in this regime.

\section{Conclusion and Discussion}		\label{section_conclusion}
In summary, we have investigated the Hamiltonian dimerization and induced quantum many-body scars in the spinless Fermi-Hubbard model with a strong-tilted field.
Interestingly, the commutation of two parts gives rise to novel oscillating behaviours and nested energy spectrum pattern, which provides a better understanding of the nonergodicity in the long-range constrained quantum systems.

\section{Acknowledgements}
This work is supported by Innovation Program for Quantum Science and Technology (Grant No. 2021ZD0301203).

\appendix

\section{Derivation for effective dimerized Hamiltonian}       \label{section_effective_model}
We start from the original Hamiltonian in Eq. (1) of main text, which can be divided into two parts $H=H_{0}+H_{1}$ with
\begin{align}
H_{0}=&\Big(-\frac{J_{\rm o}}{2}\sum_{j}c_{2j+1}^{\dag}c_{2j+2}-\frac{J_{\rm e}}{2}\sum_{j}c_{2j}^{\dag}c_{2j+1}+{\rm H.c.}\Big)  \notag \\
&+(U-\Delta)\sum_{j=1}^{L-1}(n_{j}-\frac{1}{2})(n_{j+1}-\frac{1}{2}) \,, \notag\\
H_{1}=&\sum_{j=1}^{L}\Delta j(n_{j}-\frac{1}{2})+\Delta\sum_{j=1}^{L-1}(n_{j}-\frac{1}{2})(n_{j+1}-\frac{1}{2}) \,.
\end{align}
In the rotating frame, $H_{\rm eff}=\mathcal{U}H\mathcal{U}^{\dag}-i\mathcal{U}\partial_{t}\mathcal{U}^{\dag}$ with 
$\mathcal{U}=\exp(i\int^{t}_{0}H_{1}{\rm d}t)$.
Then the effective Hamiltonian becomes
\begin{align}
H_{\rm eff}=& H_{\rm odd}+H_{\rm even}+H_{\rm off}\notag\\
=&-\frac{J_{\rm o}}{2}\sum_{j}e^{-i\Delta t}e^{i\Delta t(n_{2j}-n_{2j+3})}c_{2j+1}^{\dag}c_{2j+2} \notag \\
&-\frac{J_{\rm e}}{2}\sum_{j}e^{-i\Delta t}e^{i\Delta t(n_{2j-1}-n_{2j+2})}c_{2j}^{\dag}c_{2j+1}+{\rm H.c.}\notag\\
&+(U-\Delta)\sum_{j=1}^{L-1}(n_{j}-\frac{1}{2})(n_{j+1}-\frac{1}{2}) \,.
\end{align} 
where the $H_{\rm odd}$ and $H_{\rm even}$ denotes the constrained hopping terms acting on odd or even bonds. $H_{\rm off}$ denotes the off-resonant interaction term.
Under rotating wave approximation with $\Delta \gg J_{\rm o}, J_{\rm e}$, the first line only remains the resonant terms.
And with the condition of $U \approx \Delta$, the second line can be neglected.
Thus, the final form of effective Hamiltonian is written as
\begin{align}
H_{\rm eff}=&-\frac{J_{\rm o}}{2}\sum_{j}n_{2j}c_{2j+1}^{\dag}c_{2j+2}(1-n_{2j+3}) \notag \\
&-\frac{J_{\rm e}}{2}\sum_{j}n_{2j-1}c_{2j}^{\dag}c_{2j+1}(1-n_{2j+2})+{\rm H.c.}   \,.  
\label{sm-eq-Heff}
\end{align}

\begin{figure}[t]
\centering
\includegraphics[width=0.49\textwidth]{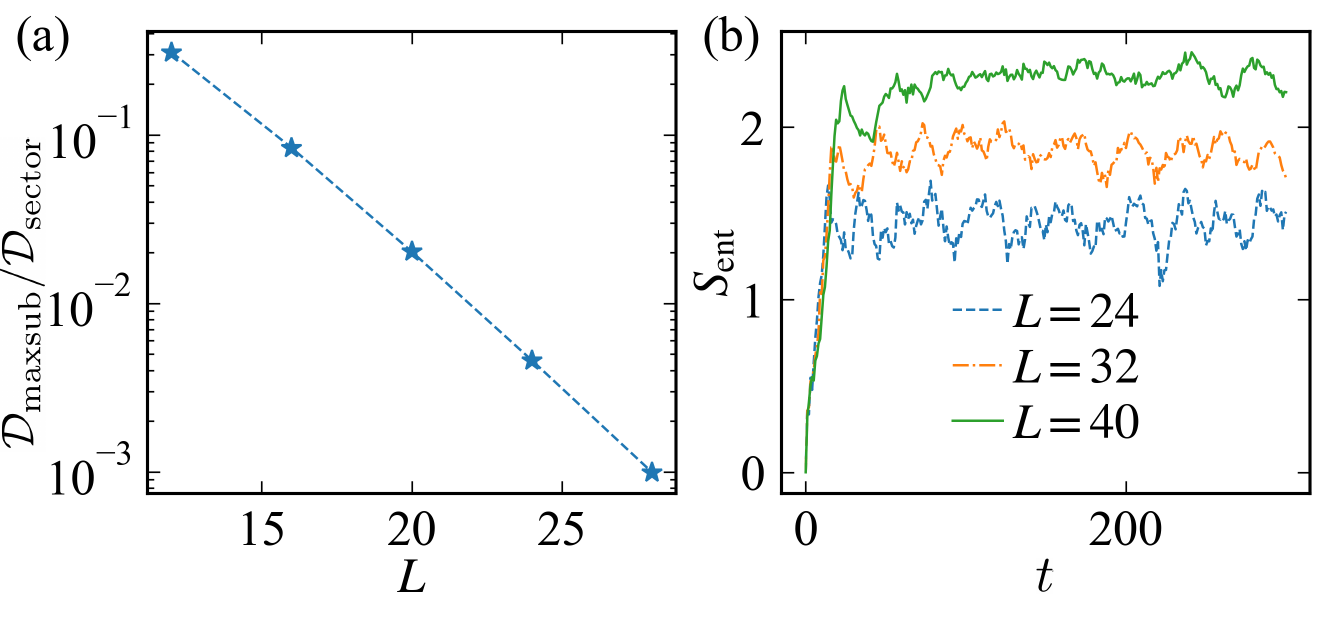}
\caption{(a) Strong Hilbert space fragmentation. Ratio between the dimension of the largest invariant subsector within corresponding $(Q, P)$ symmetry sector and the total dimension of the $(Q, P)$ sector for state $|+++\rangle$. (b) Entanglement entropy growth after a quench starting from initial state$|+++\rangle$ at $\delta =-1.0$ for different system sizes $L=24,32,40$ }\label{fig.all}
\end{figure}

\section{Mapping between tilted Fermi-Hubbard model and double PXP model}       \label{section_pxp_model}
In this section, we claim the emergence of the PXP subspace in the Fermi-Hubbard model at resonance regime 
$U \approx \Delta  \gg J_{\rm o}, J_{\rm e}$.
The PXP model is given by
\begin{align}
H=\sum_{i}^{L}P_{i}X_{i+1}P_{i+2}  \,,
\end{align}
which was first proposed in interacting Rydberg atom arrays\cite{Turner2018np}.
Here $X_{i}, Y_{i}, Z_{i}$ are the Pauli operator, $L$ denotes the system size.
The $P_{i}=(1-Z_{i})/2$ is the projector onto the the subspace spanned by configurations with no adjacent excited Rydberg atoms.

In order to understand the correspondence of Fermi-Hubbard model and PXP model,
let us first consider the revivial initial state $|1100110011001100\rangle$ 
and even part of effective Hamiltonian (\ref{sm-eq-Heff}).
Constrained by the kinetic terms, the hopping can only occur 
at two adjacent sites surrounded by the left $|1\rangle$ and right $|0\rangle$ state,
e.g., $|\cdots 1100 \cdots \rangle \rightarrow |\cdots 1010 \cdots\rangle$.
Noticing these states consist of two configurations of two-site-state $|01\rangle$ and $|10\rangle$,
the states of the PXP model can be understood to live on every two-site-state of Fermi-Hubbard chain.
The state $|01\rangle$ can be mapped to the PXP ground state $|\circ\rangle$ and $|10\rangle$ can be mapped to the PXP excited state $|\bullet\rangle$.
Using the mapping, the state $|1100110011001100\rangle\ (|1010101010101010\rangle)$ in Fermi-Hubbard model is equivalent to the state $|1\rangle\otimes|\bullet\circ\bullet\circ\bullet\circ\bullet\rangle \otimes|0\rangle\ 
(|1\rangle\otimes|\circ\circ\circ\circ\circ\circ\circ\rangle \otimes|0\rangle)$ in PXP model.
The fixed first and last site configuration of states in PXP model does not affect the evolution and statistical property \cite{Gubin2012ajp}, which can be eliminated into the periodic boundary conditions.

As for the odd part of effective Hamiltonian, it can also be mapped to the PXP model by the same method,
e.g., the state $|0110011001100110\rangle$ in Fermi-Hubbard model is equivalent to the 
$|\circ\bullet\circ\bullet\circ\bullet\circ\bullet\rangle$ state in PXP model.
Therefore, the state under both even and odd parts of Hamiltonian can be described by double PXP model jointed through fixed bond, e.g. the $|01100110110011001\rangle$ state in Fermi-Hubbard model is equivalent to the 
$|\circ\bullet\circ\bullet\rangle \otimes |1\rangle \otimes |\bullet\circ\bullet\circ\rangle$ state in double PXP model.

\section{Hilbert space fragmentation}        \label{section_fragmentation}
Compared to the conventional Hilbert space fragmentation containing dipole moment conservation\cite{Sala2020prx},
the effective Hamiltonian(\ref{sm-eq-Heff}) in tilted Fermi-Hubbard model has two U(1) conserved quantities 
$(Q, P)$ :
\begin{align}
Q=\sum_{j}n_{j} \,,	\qquad P=\sum_{j}(jn_{j}+n_{j}n_{j+1})  \,.
\end{align}
The Hilbert space organizes into small symmetry sector $\mathcal{S}$ considering these conservation $(Q,P)$.
In addition, the Hilbert space can further fractures into many disconnected subsectors $\mathcal{K}$ due to the kinetic constraint. 

Fig. \ref{fig.all}(a) shows the relation between the largest invariant subsector and corresponding $(Q, P)$ symmetry sector for state $|+++\rangle$ with the increasing system size.
Dimension of the largest invariant subspace $\mathcal{D}_{\rm max\,\,subsector}$ is exponentially smaller than dimension of the $(Q, P)$ symmetry sector $\mathcal{D}_{\rm sector}$, indicating the presence of strong Hilbert space fragmentation.
Remarkably, strong Hilbert space fragmentation leads to the absence of thermalization. 
\cite{Yang2020prl}.

\begin{figure}[t]
\centering
\includegraphics[width=0.49\textwidth]{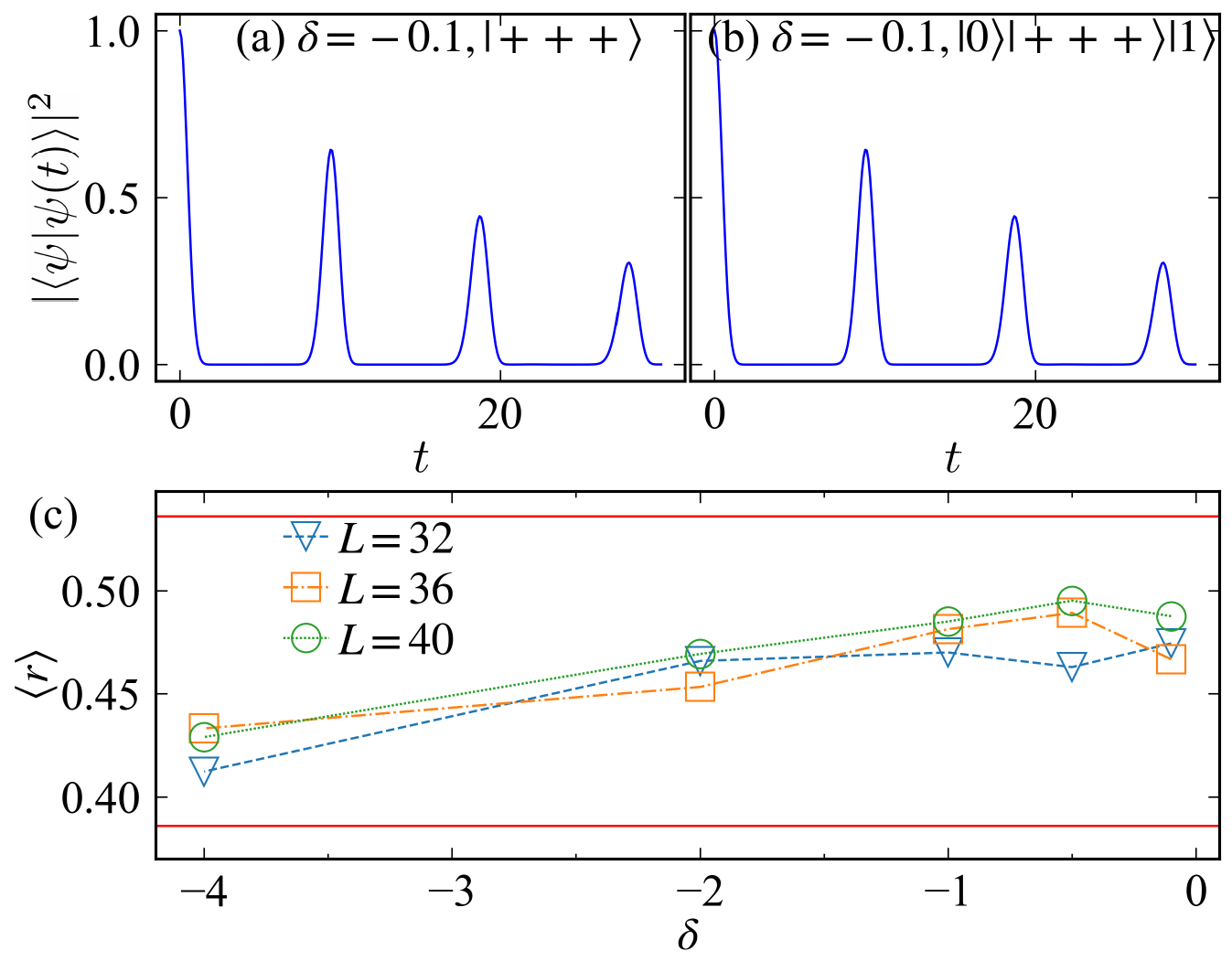}
\caption{ Fidelity dynamics for effective Hamiltonian with hopping strength $J_{\rm o}=J_{\rm e}=1$.
(a) System size $L=36$ for initial state $|+++\rangle$. 
(b) System size $L=38$ for initial state $|0\rangle\otimes|+++\rangle\otimes|1\rangle$ with the off-resonant interaction strength $\delta=-0.1$. (c) The average ratio of consecutive level spacing $\langle r\rangle$ as a function of the off-resonance strength $\delta$ for different system sizes $L=32,36,40$ respectively. The upper(below) solid line denotes  $\langle r\rangle_{w}=4-2\sqrt{3}\approx 0.536$. ($\langle r\rangle_{p}=2\ln2-1\approx 0.386$).}\label{fig.off_resonance_scar_fidelity}
\end{figure}

\section{Effects of off-resonant interaction term with initial state containing single part of effective Hamiltonian}
In this section, we show that off-resonant $(U\neq\Delta)$ interaction will cause the absence of QMBS. 
Here the effective Hamiltonian with off-resonant interaction takes the following form,
\begin{align}
H_{\rm eff}=&-\frac{J_{\rm o}}{2}\sum_{j}n_{2j}c_{2j+1}^{\dag}c_{2j+2}(1-n_{2j+3}) \notag \\
&-\frac{J_{\rm e}}{2}\sum_{j}n_{2j-1}c_{2j}^{\dag}c_{2j+1}(1-n_{2j+2})+{\rm H.c.}\notag\\
&+(U-\Delta)\sum_{j=1}^{L-1}(n_{j}-\frac{1}{2})(n_{j+1}-\frac{1}{2})   \,.
\end{align}
The off-resonance interaction term does not commute with the odd or the even part of Hamiltonian,
i.e.  $[H_{\rm off},H_{\rm odd}]\neq 0$, $[H_{\rm off},H_{\rm even}]\neq 0$, 
Consequently, under the off-resonance regime, we investigate the effect of off-resonance interaction term.

For the sake of convenience, we set fixed $\Delta=12.0$, and off-resonant interaction strength $\delta=U-\Delta$ .
Next, we investigate the effects of off-resonant interaction term through dynamical and statistical methods.
We focus on the Hilbert space containing the $|\mathbb{Z}_2\rangle$ state.
The crossover from small off-resonance strength to large off-resonance strength can be achieved by the statistical property, as shown in Fig. \ref{fig.off_resonance_scar_fidelity}(c). 
By increasing the off-resonance strength, the level statistics shows proximity from Gaussian orthogonal ensemble  to the Poissonian ensemble,
which corresponding to the phase transition from ergodicity to nonergodicity. 
We declare that it is accompanied by breakdown of QMBS.
Without loss of generality, we first focus on weak off-resonant interaction below .

\subsection{Weak off-resonant interaction}
In Fig. \ref{fig.off_resonance_scar_fidelity}, we plot the fidelity dynamics $|\langle\psi|\psi(t)$ $\rangle|^{2}$ with initial state $|+++\rangle$ and initial state $|0\rangle\otimes|+++\rangle\otimes|1\rangle$.
They both display revivals with the same period $T\approx 10$ at weak off-resonant interaction strength $\delta=-0.1$,
which means the QMBS is still existing at this regime.

\begin{figure}[t]
\centering
\includegraphics[width=0.49\textwidth]{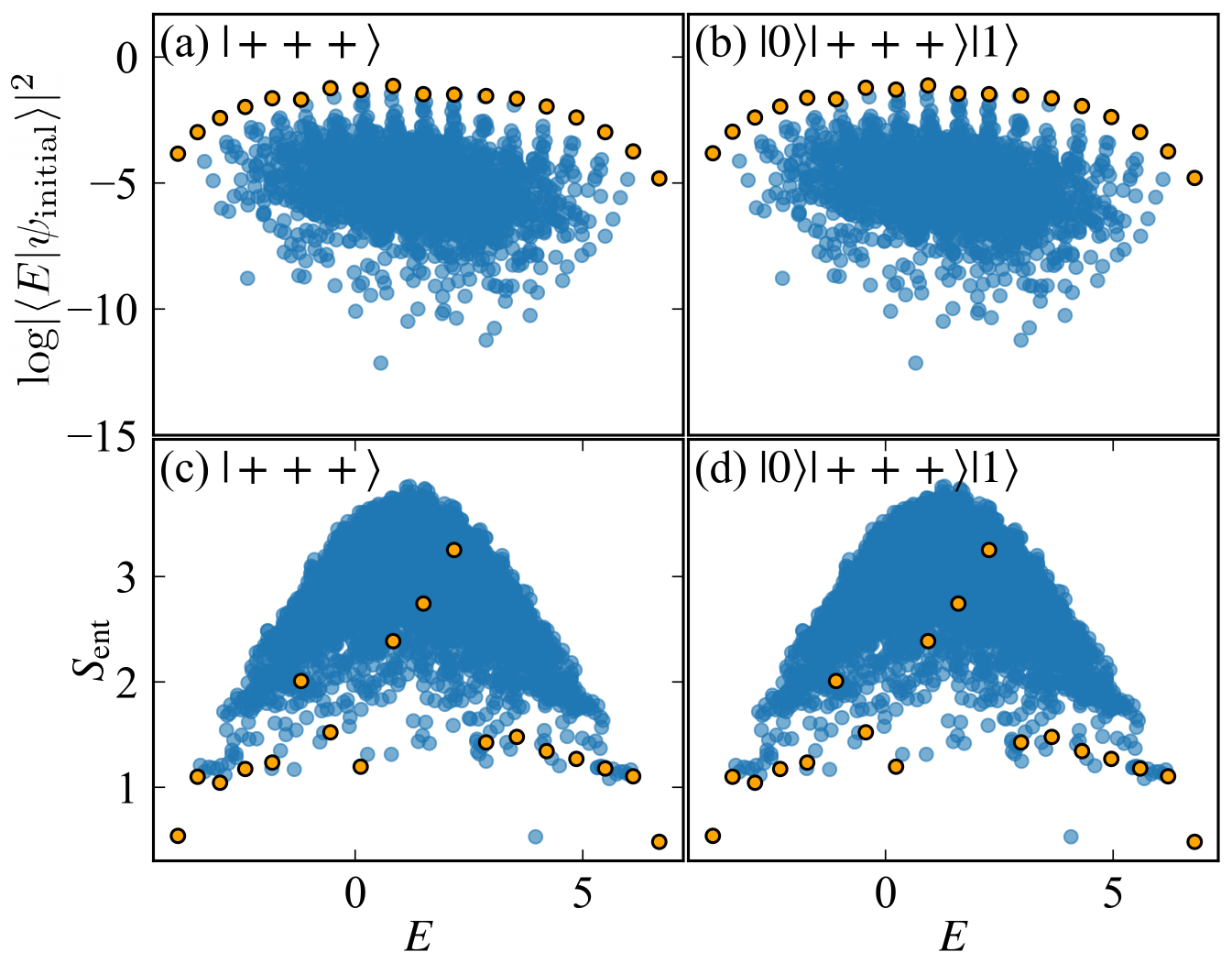}
\caption{Overlap between eigenstates and (a) $|+++\rangle$ state and  (b) $|0\rangle\otimes|+++\rangle\otimes|1\rangle$ state as a function of corresponding energy $E$.
Entanglement entropy $S_{\rm ent}$ of the eigenstates for the Hilbert space containing (c) $|+++\rangle$ state and (d) $|0\rangle\otimes|+++\rangle\otimes|1\rangle$ state.
The red dots indicate the eigenstates sitting at the top of each tower of states.
The off-resonant interaction strength $\delta=-0.1$ and hopping strength $J_{\rm o}=J_{\rm e}=1$.}\label{fig.off_resonance_entanglement}
\end{figure}

To further confirm the presence of QMBS at weak off-resonant regime.
We calculate the overlap $|\langle E|+++\rangle|^{2}$ between the different initial states ($|+++\rangle$ or $|0\rangle\otimes|+++\rangle\otimes|1\rangle$) and eigenstates $|E\rangle$ in Fig. ~\ref{fig.off_resonance_entanglement}(a-b).
It is easy to identify those special states that possess large overlap with the initial states, which are labelled by the red dots standing for the scar states .
Due to the absence of the symmetry $M=e^{i\pi\sum_{j}n_{2j+1}}$, the overlap is no longer symmetric with respect to the eigenenergy $E$.

Another evidence of existing scar can be seen in the large spread in the entanglement entropy of eigenstate in Fig. ~\ref{fig.off_resonance_entanglement}(c-d),
showing that the scar eigenstates have rather different amounts of entanglement
and less than volumn-law entanglement entropy.
Though most of the scar eigenstates still stay low entanglement, a few scar eigenstates approach high entanglement which can be originated from the off-resonant interaction.
Thus, a natural guess presents that the majority of scar eigenstates will approach high entanglement by increasing the off-resonant interaction, which leads to leakage out of QMBS.
It is noted that the properties is exactly same in both of the Hamiltonian containing the different state,
below we discuss the Hamiltonian with Hilbert space containing $|\mathbb{Z}_2\rangle$ state with the increasing off-resonant interaction.

\subsection{QMBS-Ergodicity breaking crossover with increasing off-resonant interaction}
\begin{figure}[t]
\centering
\includegraphics[width=0.49\textwidth]{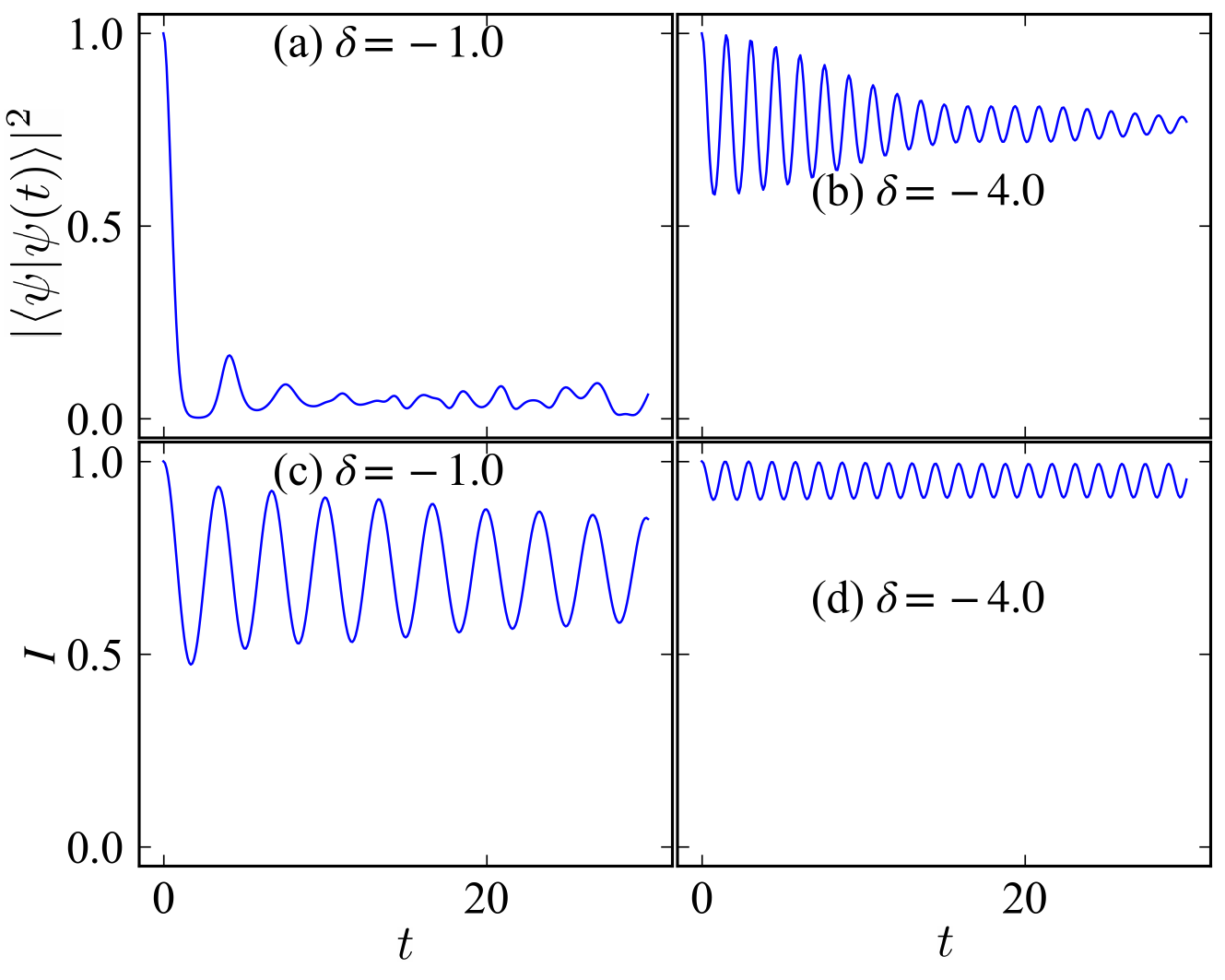}
\caption{ Fidelity dynamics for effective Hamiltonian with system size $L=36$ and hopping strength $J_{\rm o}=J_{\rm e}=1$ for initial state $|+++\rangle$ at different off-resonant interaction strengths (a) $\delta = -1.0$, (b) $\delta=-4.0$. 
The time evolution of the occupancy imbalance with system size $L=36$ and hopping strength $J_{\rm o}=J_{\rm e}=1$ for initial state $|---\rangle$ at different off-resonant interaction strengths (c) $\delta =-1.0$, (d) $\delta=-4.0$.}\label{fig.off_resonance_large}
\end{figure}

In order to distinguish whether QMBS exists in the increasing off-resonant interaction, 
we plot fidelity dynamics $|\langle\psi|\psi(t)$ $\rangle|^{2}$ at different off-resonant interaction strengths in Fig. ~\ref{fig.off_resonance_large}(a-b)(the opposite sign of off-resonant interaction can reach the same results).
Without loss of generality, we investigate the Hamiltonian with Hilbert space containing $|\mathbb{Z}_2\rangle$ state.
As the off-resonant interaction increases, the periodic oscillations vanish in Fig. ~\ref{fig.off_resonance_large}(a), which indicates the large enough off-resonant interaction breaks QMBS.
Moreover, when the off-resonant interaction reach a much larger strength than the hopping strength, system regains the periodic oscillations.
In contrast to the oscillations of QMBS, the amplitudes of oscillations are close to unity,
which shows fixed initial states indicates ergodicity breaking.

To clarify what kind of phase transition takes place during QMBS to ergodicity breaking when the off-resonant interaction strength and hopping strength are comparable,
we calculate the occupancy imbalance $\mathcal{I}=N_{\rm o}-N_{\rm e}/N_{\rm o}+N_{\rm e}$ which is a common method to further classify various quantum phases.
Fig. ~\ref{fig.off_resonance_large} (c-d) depict the dynamic evolutions of occupancy imbalance for the initial state $|---\rangle$ at different off-resonant interaction strengths.
The amplitude of the occupancy imbalance persists oscillating and stays close to the unity for a long time at $\delta =-4.0$, which suggests that fermions only slightly oscillate on fixed positions exhibiting ergodicity breaking.
The dynamic properties at the middle off-resonance strength are different from both of them.
In contrast, at a middle off-resonant interaction strength, e.g. $\delta = -1.0$,
the amplitude of the occupancy imbalance oscillates permanently and the average amplitude falls between the fixed value of thermalization $(\sim 0.47)$ and ergodicity breaking $(\sim 1)$.

In order to further investigate the properties of the intermediate phase,
we show the time evolution of half-chain entanglement entropy in Fig. \ref{fig.all}(b).
For different system sizes, entanglement entropy exhibits a linear growth first, and finally reach a saturation value, indicating a localized phase.
Compared with the many body localization systems where the entanglement scaling obeys area law\cite{Abanin2019rmp}, 
here the saturation value depends on the system size, leading to a volume-law entanglement entropy,
which suggests at first glance that the system is thermalized.
However, the entanglement entropy of thermalized systems, i.e. $S_{\rm page}\approx\frac{L}{2}\ln 2-\frac{1}{2}$,
is much higher than the saturation value of the entanglement entropy of the intermediate phase.
Therefore the intermediate phase would be a many-body localized version of the generalized Gibbs ensemble without disorder\cite{Bardarson2012prl}.

\section{Effects of off-resonant interaction term with initial states containing double part of effective Hamiltonian}	
\begin{figure}[t]
\centering
\includegraphics[width=0.49\textwidth]{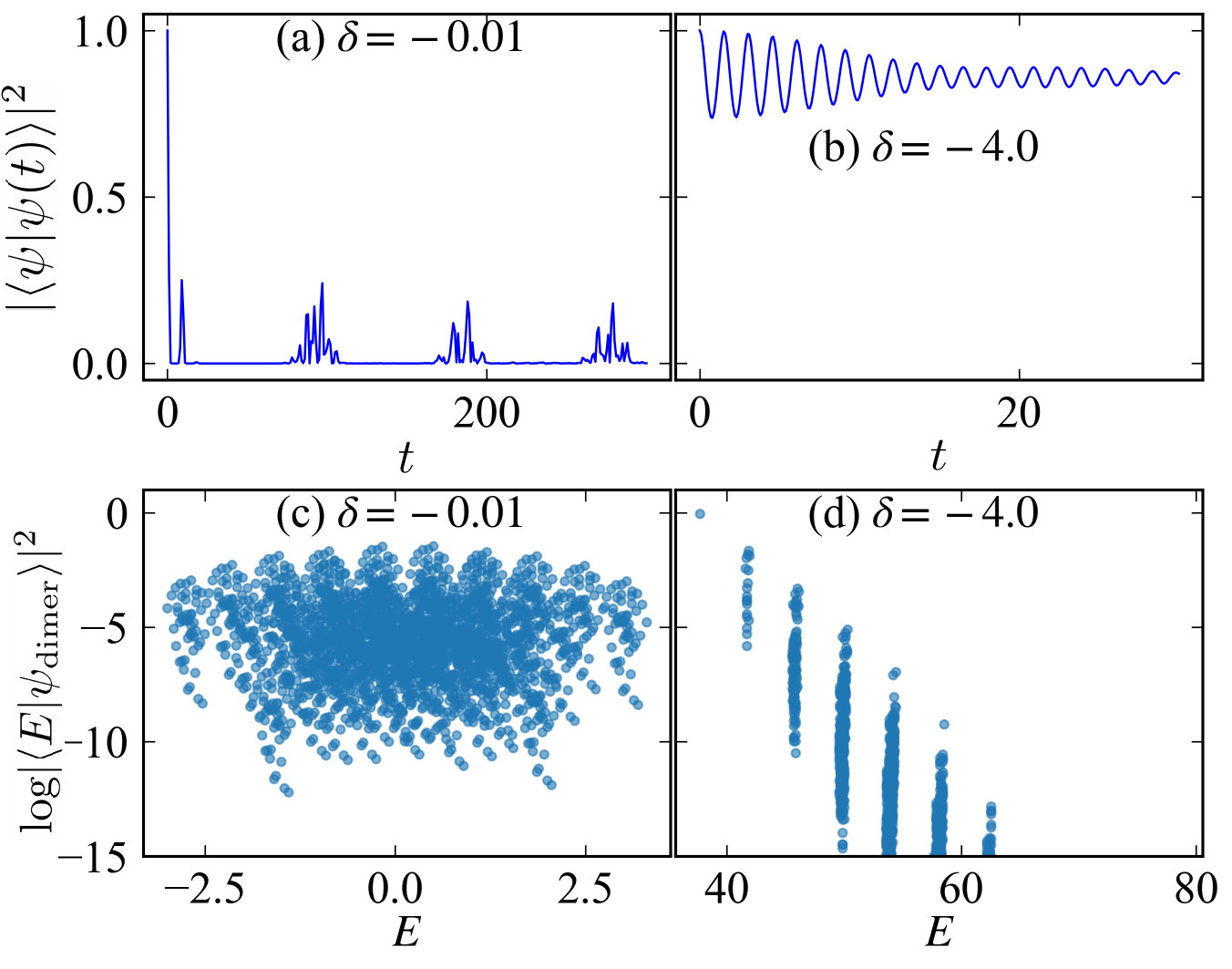}
\caption{Fidelity dynamics with $L=41$ for $|\psi_{\rm dimer}\rangle$ defined in text and $J_{\rm e}=1$, $J_{\rm o}=0.1$ at off-resonance regime (a) $\delta=-0.01$ and (b) $\delta=-4.0$. The overlap with $J_{\rm e}=1.0$, $J_{\rm o}=0.1$ (c) $\delta=-0.01$ and  (d) $\delta=-4.0$ between eigenstates and combined initial state $|\it{\Psi}_{{\rm dimer}}\rangle$ as a function of corresponding energy $E$. The system size is $L=41$.}\label{fig.off_resonance_both}
\end{figure}

In this section, we investigate the effect of off-resonant ($U\neq\Delta$) interaction with initial states built from the root configuration containing double part of effective Hamiltonian:
$|\psi_{\rm dimer}\rangle$=$|11001100\rangle\otimes |0\rangle \otimes |11001100\rangle$ ($|++0++\rangle$).
In constrast with the scar under off-resonant regime in main text, we set $J_{e}=1.0$ and $J_{\rm o}=0.1$ with $\delta=-0.01,-4.0$.

Fig. \ref{fig.off_resonance_both}(a) plots fidelity dynamics at very weak off-resonance interaction $\delta=-0.01$.
Clearly, time evolution of fidelity is same as the fidelity dynamics at the resonant regime in Fig. 3(b) in the main text.
So system reserves the properties of nested scar at tiny off-resonance interaction,
which is consistent with the phenomena in the single part of Hamiltonian.
The corresponding overlap between eigenstates and combined initial state $|\psi_{\rm dimer}\rangle$ present a set of tower,
consisting of additional equal points in Fig. \ref{fig.off_resonance_both}(c).

The fidelity dynamics at strong off-resonance interaction $\delta=-4.0$ is shown in Fig. \ref{fig.off_resonance_both}(b).
After a long period of evolution, system fidelity stays 0.5 which is large than the usual scar states.
It means that strong off-resonance interaction may break the QMBS.
At the same time, nonzero fidelity indicates the system maintains the initial memory,
leading to the breaking of ergodicity.
Surprisingly, the overlap at off-resonance interaction $\delta=-4.0$ presents individual aggregation distribution at fixed values in Fig. \ref{fig.off_resonance_both}(d).
This phenomenon is based on the strength of off-resonance interaction is much stronger than the stronger of hopping terms.
Different configuration states may cause different energy of interaction,
in which some of configurations share the same interaction energy and energy difference is strength of off-resonance interaction $\delta$.
Based on the different interaction energy, the hopping is the tiny disturbance.
It means the states with same interaction energy are almost degenerate, leading to additional symmetry.
The additional symmetry cause the further fragmentation based on the subspace $\mathcal{K}$ and the ergodicity breaking is further aggravated by the extra symmetry.
The explanation about the ergodicity breaking will be more reasonable with the stronger off-resonance strength.

\bibliography{ref}
\end{document}